\documentclass[fleqn,usenatbib]{mnras}

\usepackage{newtxtext,newtxmath}
\usepackage[T1]{fontenc}
\usepackage{ae,aecompl}

\usepackage{graphicx}	
\usepackage{amsmath}	
\usepackage{amssymb}	
\usepackage{atbegshi}

\title{$^{12}$C/$^{13}$C isotopic ratios in red-giant stars of the open cluster NGC~6791}

\author[Szigeti et al.]{
L{\'a}szl{\'o} Szigeti$^{1}$\thanks{E-mail: szilac@gothard.hu},
Szabolcs~M{\'e}sz{\'a}ros$^{1,2}$,
Verne~V.~Smith$^{3}$,
Katia Cunha$^{4,5}$,
\newauthor
Nad{\`e}ge Lagarde$^{6}$,
Corinne Charbonnel$^{7,8}$,
D.~A.~Garc\'{\i}a-Hern{\'a}ndez$^{9,10}$,
\newauthor
Matthew Shetrone$^{11}$,
Marc Pinsonneault$^{12}$,
Carlos Allende Prieto$^{9,10}$,
\newauthor
J. G. Fern{\'a}ndez-Trincado$^{13,14,15}$,
J{\'o}zsef Kov{\'a}cs$^{1}$,
Sandro Villanova$^{13}$
\\
$^{1}$ELTE E\"otv\"os Lor\'and University, Gothard Astrophysical Observatory, Szombathely, Hungary\\
$^{2}$Premium Postdoctoral Fellow of the Hungarian Academy of Sciences\\
$^{3}$National Optical Astronomy Observatory, Tucson, AZ 85719, USA\\
$^{4}$Steward Observatory, University of Arizona, Tucson, AZ 85719\\
$^{5}$Observat\'orio Nacional/MCTI, Rio de Janeiro, Brazil\\
$^{6}$Institut Utinam, CNRS UMR6213, Univ. Bourgogne Franche-Comt{\'e}, OSU THETA Franche-Comte Bourgogne,\\ 
$^{7}$Department of Astronomy, University of Geneva, Chemin des Maillettes 51, 1290 Versoix, Switzerland\\
$^{8}$IRAP, UMR 5277, CNRS and Universit{\'e} de Toulouse, 14, av.E.Belin, 31400 Toulouse, France\\
Observatoire de Besan\c{c}on, BP 1615, 25010 Besan\c{c}on Cedex, France\\
$^{9}$Instituto de Astrof{\'{\i}}sica de Canarias (IAC), E-38205 La Laguna, Tenerife, Spain\\
$^{10}$Universidad de La Laguna, Departamento de Astrof\'{\i}sica, 38206 La Laguna, Tenerife, Spain\\
$^{11}$University of Texas at Austin, McDonald Observatory, USA\\
$^{12}$Department of Astronomy, The Ohio State University, 140 W. 18th Ave., Columbus, OH 43210, USA\\
$^{13}$Departmento de Astronom{\'i}a, Universidad de Concepc{\'i}on, Casilla 160-C, Concepc{\'i}on, Chile\\
$^{14}$Departamento de Astronom\'\i a, Casilla 160-C, Universidad de Concepci\'on, Concepci\'on, Chile\\
$^{15}$Institut Utinam, CNRS UMR6213, Univ. Bourgogne Franche-Comt\'e, OSU THETA , Observatoire de Besan\c{c}on,\\ 
BP 1615, 25010 Besan\c{c}on Cedex, France\\
}

\date{Accepted XXX. Received YYY; in original form ZZZ}

\pubyear{2017}

\begin{document}
\label{firstpage}
\pagerange{\pageref{firstpage}--\pageref{lastpage}}
\maketitle

\begin{abstract}
Carbon isotope ratios, along with carbon and nitrogen abundances, are derived in a sample of 11 red-giant members of one of the most metal-rich clusters in the Milky Way, NGC~6791. The selected red-giants have a mean metallicity and standard deviation of [Fe/H]=+0.39$\pm$0.06 (Cunha et al. 2015).  
We used high resolution H-band spectra obtained by the SDSS-IV Apache Point Observatory Galactic Evolution Experiment (APOGEE). The advantage of using high-resolution spectra in the H-band is that lines of CO are well represented and their line profiles are sensitive to the variation of $^{12}$C/$^{13}$C.
Values of the $^{12}$C/$^{13}$C ratio were obtained from a spectrum synthesis analysis.  
The derived $^{12}$C/$^{13}$C ratios varied between 6.3 and 10.6 
in NGC~6791, in agreement with the final isotopic ratios from thermohaline-induced mixing models. 
The ratios derived here are combined with those obtained for more metal poor red-giants from the literature to examine the correlation between $^{12}$C/$^{13}$C, mass, metallicity and evolutionary status. 
\end{abstract}




\section{Introduction}

Over the last decade, large surveys (``big data'') in astronomy have advanced significantly our understanding of the structure of the Milky Way. The currently 
operating high-resolution spectroscopic surveys, Gaia-ESO \citep{Gilmore2012}, GALAH \citep{Martell2017},  
APOGEE \citep{Majewski2017, Blanton2017}, and low-resolution survey, LAMOST \citep{Xiang2017} are mapping the chemical composition of the Milky Way using different wavelength regions and somewhat different spectral resolutions. 

The Apache Point Observatory Galactic Evolution Experiment (APOGEE) was one component of the $3^{\rm rd}$ phase of the Sloan Digital 
Sky Survey (SDSS-III; \citep{Eisenst}) and continues as part of SDSS$-$IV. The goal of APOGEE is to obtain high-resolution (R = 22500), 
high signal-to-noise, H-band spectra ($\lambda$ = 1.51$-$1.69$\mu$m) of nearly 500,000 (mostly) red-giant stars in the Milky Way by the end of 2020, 
and to determine chemical abundances of as many as 23 elements in these stars\footnote{http://www.sdss.org/dr13/irspec/}. 
Majority of APOGEE targets are evolved red-giant branch (RGB), red clump (RC), or
asymptotic giant branch (AGB) stars from all major Galactic stellar populations. 
APOGEE is unique amongst the current high-resolution spectroscopic surveys because it observes in the near infrared and uses telescopes \citep{Bowen1973, Gunn2006} in 
both the Northern and Southern hemispheres\footnote{http://www.sdss.org/surveys/apogee-2/}.
The APOGEE data reduction Pipeline (ASPCAP) \citep{Zamora2015, Nidever2015, ASPCAP_tech} produces one-dimensional, 
wavelength and flux calibrated spectra, corrected for terrestrial airglow lines, and 
telluric absorption lines.  The reduced spectra are then analysed via the APOGEE Stellar Parameters and Chemical Abundance Pipeline \citep{ASPCAP_tech}
to determine both 
fundamental stellar parameters, as well as detailed chemical abundances.  
The current version
of ASPCAP does not measure $^{12}$C/$^{13}$C ratios but these are, measurable from CO and CN lines in the observed spectral
window, and carry important information about stellar nucleosynthesis
and mixing processes along the RGB and AGB. 
Future versions of ASPCAP are being modified to derive carbon
isotope ratios; however, in this study manual determinations of
$^{12}$C/$^{13}$C ratios are presented for RGB and RC 
members of the old,
metal-rich cluster NGC~6791, which are used to both probe 
stellar evolution, nucleosynthesis, and mixing, as well as
providing results which will be used to provide checks for future ASPCAP-derived 
values of $^{12}$C/$^{13}$C.

During stellar evolution on the main sequence, the surface carbon 
isotopic ratio represents the composition of the interstellar cloud 
from which the star formed, which is high, e.g.,
$^{12}$C/$^{13}$C= 89 for the Sun \citep{Asplund2009}. 
When a low-mass star evolves to the base of the RGB, the outer 
convective envelope reaches its greatest extent in mass and penetrates 
into layers where the chemical composition has been partially altered 
by H-burning on the CN cycle. Convection brings this material to the 
stellar surface, where the effects of CN-cycle proton-captures can be
observed spectroscopically, e.g., a somewhat lower ratio of C/N, along
with a significantly lower ratio of $^{12}$C/$^{13}$C.
This phase of stellar evolution is referred to as the first 
dredge-up \citep{Iben1965}.  Standard stellar evolution models for 
stars with a mass near that of the Sun predict that the 
$^{12}$C/$^{13}$C ratio decreases to a value of $\sim$29 (from an
initial value of 89) \citep{Charbonnel1994}.  Carbon and Nitrogen
surface abundances are not predicted to change after the first dredge-
up in standard model. However, observations 
show that this is not the case \citep[e.g.][]{Charbonnel1998, Gratton2000, Shetrone2003, Recio2007, Mikolaitis2012, Tautvaisiene2015, Tautvaisiene2016}: the abundance of 
carbon decreases and the abundance of nitrogen increases further as the star climbs the RGB. In order to explain the observations, one requires 
non-canonical extra mixing. Several mechanisms have been suggested \citep{Tautvaisiene2016}[and references therein], but 
in recent years thermohaline mixing \citep{Charbonnel2010} has emerged as a leading theory that can explain many of the observations.

Thermohaline mixing was first used in a stellar context by \citet{Stothers1969}. 
In red giants, the instability is triggered by the molecular weight inversion \citep{Ulrich1972}, created by the $^3 \rm He(^3 \rm He,2p)^4 \rm He$ 
reaction \citep{CharbonnelandZahn2007}, when the star reaches the so called luminosity bump \citep{Iben1968, King1985}. 
The reaction starts in the external wing of the hydrogen burning 
shell when the shell crosses the molecular weight barrier left behind by the $1^{\rm st}$ 
dredge up. When the $^3 \rm He$ begins to burn, the local mean molecular weight decreases and the temperature becomes higher 
than its surroundings. Due to the change from two initial to three final particles, the cell expands to establish pressure equilibrium. 
The expansion reduces the density and the whole cell rises until the external density and pressure become equal to the internal state \citep{Charbonnel2010}. There will also be a 
thermal gradient between the bubble and the surroundings.
When the temperature and molecular weight gradient build up, they trigger the mixing. Heat begins to diffuse 
into the bubble faster than the particles, which takes the form of "long fingers". These fingers sink because they are heavier 
than the environment until they become turbulent and dissolve.
Thermohaline mixing is one possible source of changes in surface abundances observed
in Li, $^{12}$C, $^{13}$C, and $^{14}$N among luminous RGB stars.
The prescription for this mechanism used in current stellar models are however very simplistic. They are currently being challenged by hydrodynamical 2D and 3D simulations which are still far from reaching the actual conditions relevant to stellar interiors \citep[e.g.][]{Traxler2011, Brown2013, Wachlin2014} In this context, it is crucial to determine the abundances of the above mentioned chemical elements in homogeneous stellar samples with good evolution constraints as is the case in open clusters.  

Previous studies in the literature determined $^{12}$C/$^{13}$C ratios in stars close to solar metallicity
\citep[e.g.][]{Gilroy1989, Gilroy1991, Tautvaisiene2000, Shetrone2003_a, Tautvaisiene2005, Smiljanic2009, Mikolaitis2010, Mikolaitis2011_a, Mikolaitis2011_b, Mikolaitis2012, Santric2013,  Tautvaisiene2016, Drazdauskas2016, Dradauskas2016_b}. 
Parts of the H-band spectra are sensitive to the variation of $^{12}$C/$^{13}$C ratio, and the APOGEE survey allows us to derive its value for a broad metallicity range above [Fe/H] $>-$0.6. The most metal-rich stars previously observed were in NGC~6253 \citep{Mikolaitis2012} with metallicities of [Fe/H] $\sim$ +0.46 dex.
We derived  $^{12}$C/$^{13}$C ratios from spectra of the most metal-rich stars available in the APOGEE data base, those in NGC~6791, in order to investigate the behavior of the carbon isotopic ratio in the atmospheres of high-metallicity stars.  


\section{Data and Methods}

\subsection{Targets Analysed}

Red-giants in NGC~6791, one of the most metal-rich open clusters in the Milky Way, are ideal targets for the examination of extra mixing in a high-metallicity regime.
This open cluster has a mean metallicity of [Fe/H] = +0.39 (A(Fe)$_{\rm Sun}$=7.45), as derived in the previous APOGEE study of red-giants from a manual abundance analysis of sodium and oxygen by \citep{Cunha2015}. Recent studies \citep[e.g.][]{Ness2017, Linden2017} have different metallicities but we used the value from \citet{Cunha2015} because of the consistency. 
This cluster has an age of $\sim$8 Gyr and it is the youngest one in the $\alpha$-rich thick disc or bulge population \citep{Linden2017}.  

In this study, we analyse the same 11 evolved stars selected by \citet{Cunha2015}; their membership  has been discussed in \citet{Frinchaboy2013, Cunha2015} and also a radial-velocity membership study \citep{Tofflemire2014} indicates that all of these targets are cluster members. All of the observed spectra have high ( $>90$ ) signal-to-noise ratios in order to minimize random observational error. 

The basic stellar parameters and metallicities ($T_{\rm eff}$, log~g, [Fe/H]) for the target stars are listed in Table 1. A detailed description of the atmospheric parameter determinations can be found in \citet{Cunha2015}. 

\begin{table*}                                                                                                                                     
\begin{tabular}{lccccccccc}                                                                                           
\hline                                                                                                                
2MASS ID & S/N & v$_{\rm helio}$ & T$_{\rm eff}$ & $\log~g$ & [Fe/H] & [C/Fe] & [N/Fe] & $\xi$ & Evol. \\              
 &  & [$\frac{km}{s}$] & [K] & & & & & [$\frac{km}{s}$] & state \\                                                    
\hline                                                                                                                
J19204557+3739509 &   92  & -46.3 & 4500 & 2.45 & 0.44 & -0.09  & 0.41 & 1.3	&  RC			\\                        
J19204971+3743426 &   368 & -47.5 & 3530 & 0.80 & 0.40 & -0.19  & 0.45 & 1.8	&	...			\\                        
J19205338+3748282 &   154 & -48.5 & 4075 & 1.70 & 0.42 & -0.11  & 0.31 & 1.7	&  RGB			\\                        
J19205510+3747162 &   159 & -48.2 & 4000 & 1.62 & 0.38 & -0.05  & 0.36 & 1.7	&	RGB		\\                        
J19205530+3743152 &   113 & -47.9 & 4300 & 2.27 & 0.42 & -0.04  & 0.36 & 1.7	&	RGB			\\                        
J19210112+3742134 &   135 & -47.7 & 4255 & 2.21 & 0.33 &  0.00  & 0.47 & 1.6	&	RGB			\\                        
J19210426+3747187 &   132 & -46.8 & 4200 & 1.85 & 0.40 & -0.10  & 0.35 & 1.8	&	RGB		\\                        
J19210483+3741036 &   98  & -50.0 & 4490 & 2.48 & 0.39 & -0.15  & 0.36 & 1.5	&	RC			\\                        
J19211007+3750008 &   103 & -49.0 & 4435 & 2.66 & 0.41 & -0.06  & 0.37 & 1.3	&	RGB			\\                        
J19211606+3746462 &   856 & -46.8 & 3575 & 0.76 & 0.29 & -0.17  & 0.46 & 1.8	&	...			\\                       
J19213390+3750202 &   552 & -47.0 & 3800 & 1.25 & 0.28 & -0.09  & 0.40 & 1.7	&	...			\\                        
\hline 
\end{tabular}  
\caption{Atmospheric parameters of selected stars in NGC~6791. The evolutionary state determination is from the APOKASC catalog \citep{Pinsonneault2014}}
\end{table*}

\subsection{Abundance Analysis}

Carbon and nitrogen abundances were not published by \citet{Cunha2015}. Molecular lines of CO were used to derive carbon abundances, CN lines for deriving nitrogen abundances, and OH lines were used for obtaining oxygen abundances (see \citet{Cunha2015} for details). The methodology described in \citet{Verne2013} was used to derive abundances of C, N and O by fitting of all molecular lines consistently. 
The selected line list adopted in the calculations of synthetic spectra was the same used for DR13 and described in detail by \citet{linelist}. 	

The current version of ASPCAP (DR13 \citep{DR13}) does not determine the $^{12}$C/$^{13}$C ratio, thus we chose an independent code called \textit{autosynth} \citep{autosynth}, which compares theoretical spectra with observations and 
determines abundances via a $\chi^2$ minimization. 
Our synthetic spectra were based on 1D Local Thermodynamic Equilibrium (LTE) model atmospheres calculated with ATLAS9 \citep{kurucz1979, kurucz1993} using the solar mixture from \citet{Asplund2005}. 
Scaled with the metallicity, model atmospheres were computed using the method described in \citet{Meszaros2012}. 

For spectrum syntheses, the \textit{autosynth} gives the parameters to the MOOG2013\footnotetext{http://www.as.utexas.edu/~chris/moog.html} \citep{Sneden1973} and
MOOG creates a synthetic spectrum using model atmospheres calculated by adopting atmospheric parameters such as, T$_{\rm eff}$, log~g, [Fe/H] and abundances of [C/Fe], [N/Fe] taken from \citet{Cunha2015}.
Afterwards, this model spectrum is then compared to the observations.

\begin{figure*}
\centering
\includegraphics[width=4.6in,angle=270]{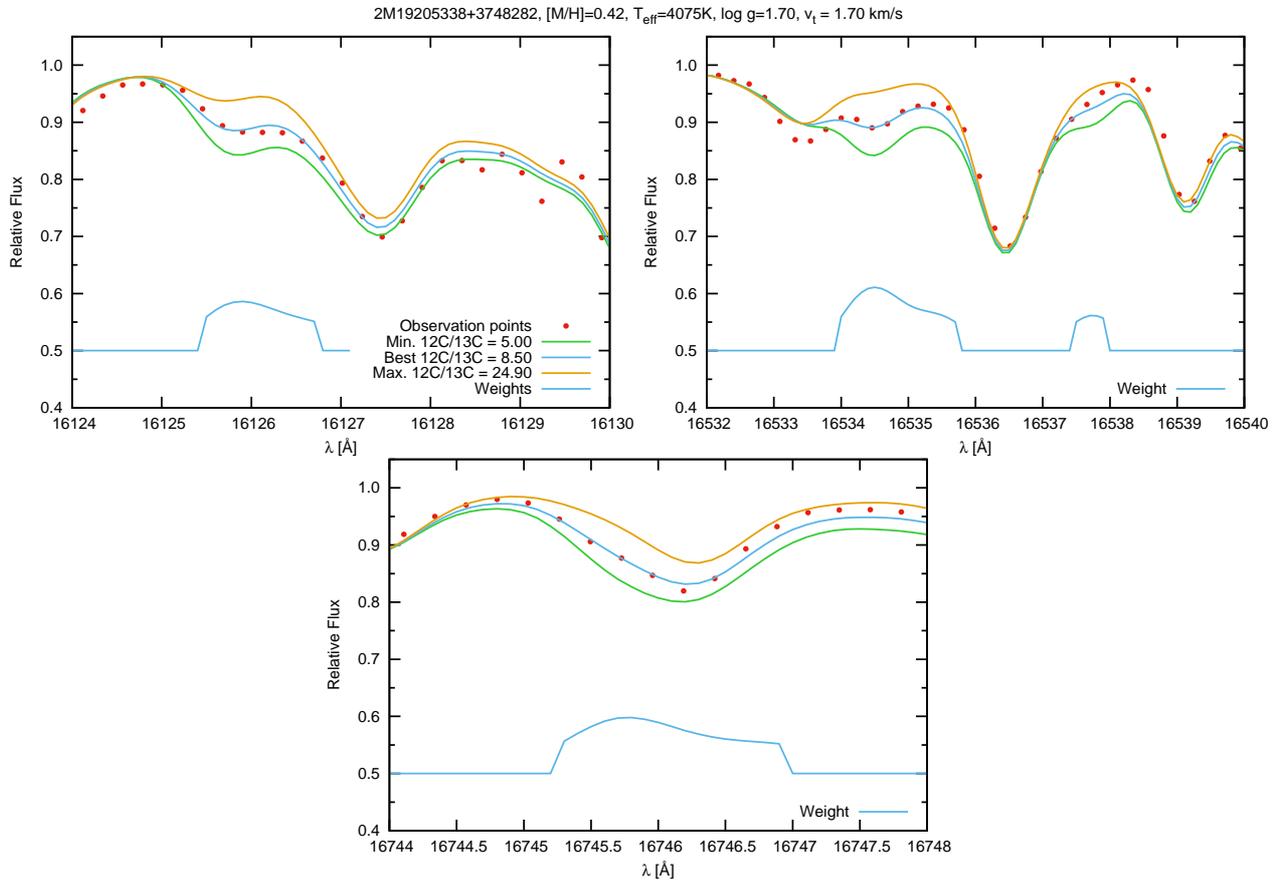}
\caption{Examples of our selected wavelength regions. $^{12}$C/$^{13}$C ratios have been fitted between 
the labelled boundaries denoted by orange and green lines, while the best-fitting is denoted by the blue line. 
} 
\label{fig:spectra1} 
\end{figure*}

After feeding the main atmospheric parameters and abundances into the program, we fitted the $^{12}$C/$^{13}$C ratios in pre-defined wavelength windows centred around $^{12}$CO and $^{13}$CO molecular lines. 
These windows were determined by subtracting two synthetic spectra, one with $^{12}$C/$^{13}$C=100, and one with $^{12}$C/$^{13}$C=2, from each other. 
The resulting difference highlights the spectral regions most sensitive to the $^{12}$C/$^{13}$C ratio. Only those segments of the spectrum were chosen, where the difference between the two normalized spectra was higher than 0.025. This difference was used as weights during the fitting process, in which the value of $^{12}$C/$^{13}$C was varied between 5 and 25 for all stars. Examples of wavelength windows are 
shown in Figure \ref{fig:spectra1}. The selected regions were: 16125.5 $-$ 16126.7 \AA, 16534.0 $-$ 16535.9 \AA, 16537.5 $-$ 16537.9 \AA~ and 16745.3 $-$ 16746.9 \AA. The resulting isotopic ratios are listed in Table 2. The second window represent both the second and the third wavelength regions.  

The uncertainties in the atmospheric parameters, T$_{\rm eff}$, log~g and [Fe/H], affect the abundances, and therefore, the final $^{12}$C/$^{13}$C ratios as well. The uncertainties were calculated by changing the T$_{\rm eff}$ by 100K, [Fe/H] by 0.1~dex, log~g by 0.2~dex one-by-one (these values generally represent the average ASPCAP errors for this cluster), and then by repeating the same fitting procedure described in the previous section. 
The individual error from each atmospheric parameter was the difference between the original $^{12}$C/$^{13}$C and these altered calculation values. The final value of uncertainty was the sum in quadrature of these three individual errors. Asymmetric values of the errors were calculated because of the carbon isotopic ratio reciprocal nature. 
We would like to acknowledge that we are taking shortcuts to computing errors. Firstly, we are examining how one value propagates, rather than sampling the error distribution in the parameters and secondly, we are assuming errors in the atmospheric parameters are uncorrelated.

\begin{table*}
\caption{The results and the errors}
\begin{tabular}{lrrrccccc}
\hline
2MASS ID & $T_{eff}$ & $^{12}C/^{13}C$ & $^{12}C/^{13}C$ & $^{12}C/^{13}C$ & $^{12}C/^{13}C$ & Error$+$ & Error$-$ \\
 & [K] & window 1.$^a$ & window 2.$^b$ & window 3.$^c$ & average$^d$ & & \\
\hline
J19204557+3739509  &  4500 &  8.80 & 5.80 & 5.20  & 6.3     &   1  &   1  \\
J19204971+3743426  &  3530 &  ...  & 10.6 & 10.5  & 10.6    &   2  &   2  \\
J19205338+3748282  &  4075 &  8.50 & 8.60 & 8.40  & 8.5     &   1  &   2  \\
J19205510+3747162  &  4000 &  ...  & 10.5 & 8.50  & 9.4     &   1  &   2  \\
J19205530+3743152  &  4300 &  ...  & 11.0 & 8.80  & 9.8     &   2  &   3  \\
J19210112+3742134  &  4255 &  ...  & 14.1 & 8.50  & 10.6    &   2  &   3  \\
J19210426+3747187  &  4200 &  10.4 & 9.20 & 7.10  & 8.7     &   2  &   2  \\
J19210483+3741036  &  4490 &  7.3  & 8.6  & ...   & 7.95    &   2  &   4  \\
J19211007+3750008  &  4435 &  ...  & 6.60 & 6.90  & 6.8     &   1  &   1  \\
J19211606+3746462  &  3575 &  11.2 & 8.10 & 9.0   & 9.3     &   2  &   3  \\
J19213390+3750202  &  3800 &  ...  & 11.4 & 9.20  & 10.2    &   2  &   2  \\
\hline
\multicolumn{3}{l}{\footnotesize$^a$ Fitting range: 16120 \AA~ - 16133 \AA}\\
\multicolumn{3}{l}{\footnotesize$^b$ Fitting range: 16530 \AA~ - 16545 \AA}\\
\multicolumn{3}{l}{\footnotesize$^c$ Fitting range: 16738 \AA~ - 16754 \AA}\\
\multicolumn{3}{l}{\footnotesize$^d$ Mean of the three distinct value}\\
\end{tabular}
\end{table*} 


\section{Discussion}

\subsection{Carbon Isotope ratios and Models in NGC~6791}

The derived values of the carbon isotopic ratios in NGC~6791 are lower than expected from the standard 1$^{\rm st}$ dredge-up model, typically around 26$-$30 depending on mass.
An extra mixing mechanism is presumably responsible for this discrepancy in the surface carbon ratio. 
\citet{Shetrone2003_a} derived the carbon isotopic ratio for 32 stars in 4 globular cluster and they pointed out, that the 
extra mixing effect occurs above the luminosity bump and this is what the thermohaline mixing model predicts. See also \citet{Charbonnel1998}													

\begin{figure*}
\centering
\includegraphics[width=4in,angle=270]{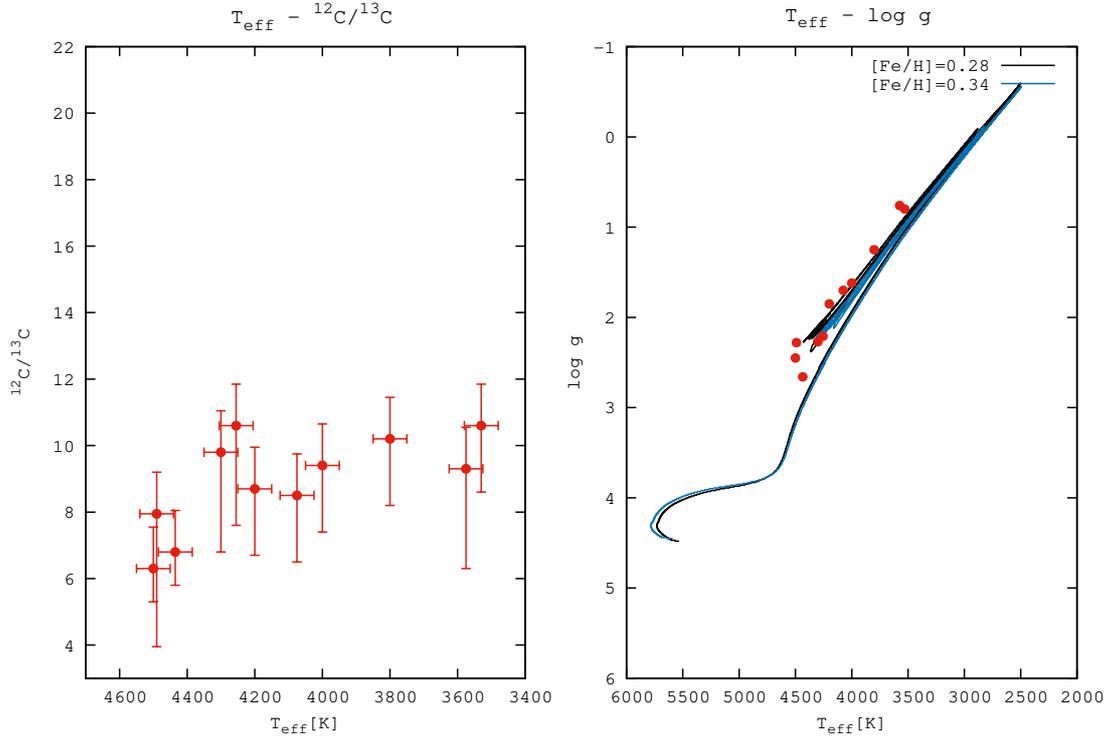}
\caption{Left panel: $^{12}$C/$^{13}$C ratio as a function of effective temperature from this paper. Right panel: Log~g as 
a function of effective temperature with evolutionary tracks from \citet{Lagarde2017} at different metallicities. 
}
\label{fig:TCratio}
\end{figure*}

From Figure \ref{fig:TCratio} (left panel), we can see that the $^{12}$C/$^{13}$C ratio in our sample stars does not correlate significantly with the effective temperature, although the three hottest red giants exhibit somewhat lower ratios and these stars are most red clump giants. The amount of extra mixing visible in NGC~6791 could reduce the isotopic ratio to between 6.3 and 10.6, and this is in good correspondence with the lowest value obtained at the RGB tip when thermohaline mixing is taken into account \citep{Charbonnel2010}.
This study indicates that extra mixing occurs at very high metallicities, similarly to solar-metallicity stars.

As mentioned previously, the carbon isotopic ratios undergo drastic changes in the RGB. 
Stellar models have been computed for the present paper using the stellar evolution code STAREVOL \citep[e.g.][]{Lagarde2012} for two stellar masses (1.15 and 1.10 M$_\odot$) at two metallicities ([Fe/H]=0.34 and 0.28). We use the same main physical ingredients that are used and described in \citet{Lagarde2017}. To quantify the effects of thermohaline mixing we computed stellar evolution models assuming the standard prescription (no mixing mechanism other than convection), and including the effects of thermohaline instability induced by $^3$He burning as described by \citet{Charbonnel2010}.

\begin{figure*}                                                                                                 
\centering                       
\includegraphics[width=4in,angle=270]{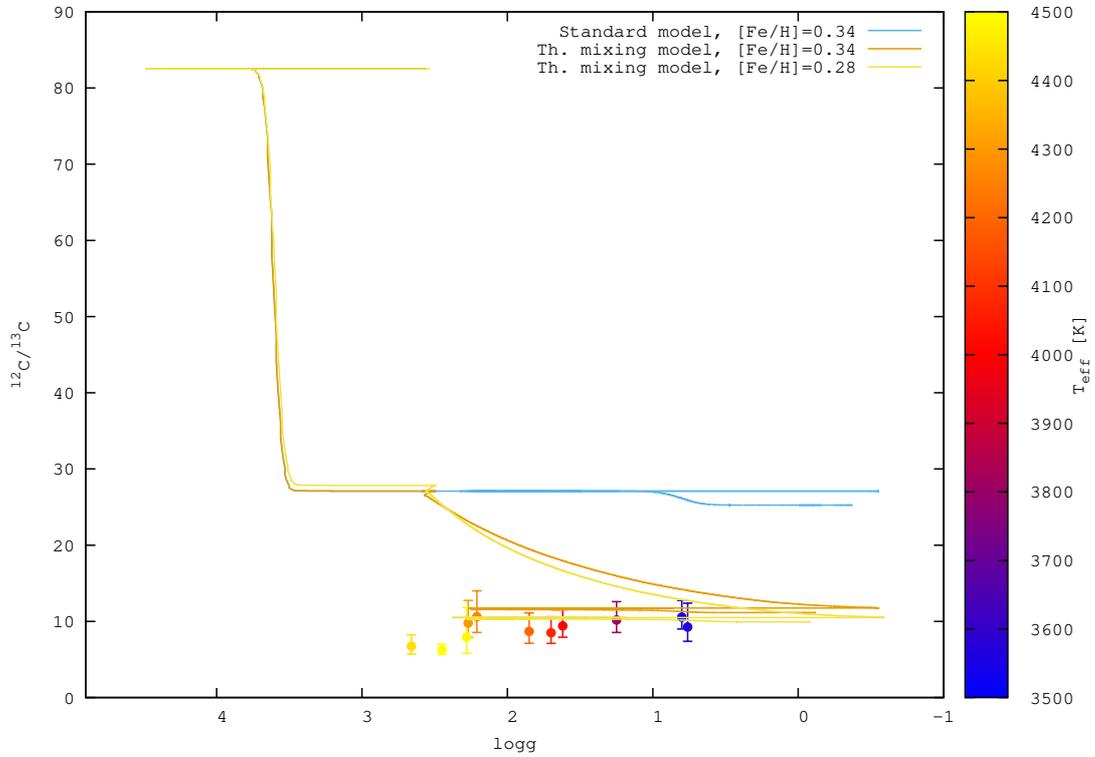}     
\caption{
The evolution of the theoretical surface carbon isotopic ratio computed with STAREVOL following the standard prescription for 1.15$\rm M_{\odot}$ model at [Fe/H] =0.34 (blue solid line), and including the effects of thermohaline instability for $1.1\rm M_{\odot}$ model at [Fe/H]=0.34 (orange solid line) and for 1.10$\rm M_{\odot}$ model at [Fe/H]=0.28 (yellow solid line). Observations are compared with models and color-coded according to the effective temperature.}                                                                                                
\label{fig:c-logg_model}                                                                                                                     
\end{figure*}

In Figure \ref{fig:c-logg_model} we plot the theoretical predictions of our stellar models including thermohaline mixing and computed at the average metallicity of NGC~6791 and compare with our observations. The carbon isotopic ratio first decreases under the effect of the first dredge-up (log~g $\sim$ 3.7).  The value of $^{12}$C/$^{13}$C decreases further after the RGB bump (log~g $\sim$ 2.5), as the star evolves up the RGB to the tip.
After the He core-flash, the carbon isotope ratio remains fairly constant during the RC and early-AGB phases of evolution.  

The final, asymptotic values of $^{12}$C/$^{13}$C predicted by the thermohaline mixing 
models, shown in Figure 3, agree
very well with the derived carbon isotopic ratios in the NGC 6791 red giants. However, an inconsistency remains, in the behavior of the carbon isotope ratio as a function of
log g (which is a proxy for luminosity).  Instead of a gradual decline in
$^{12}$C/$^{13}$C with increasing luminosity (or decreasing log g) above the luminosity
bump, as predicted by the thermohaline model, the observed carbon isotope ratios
display a rapid decrease in $^{12}$C/$^{13}$C right after the luminosity bump. 
The predicted behavior of $^{12}$C/$^{13}$C as a function of RGB luminosity from
thermohaline mixing models can depend on the detailed numerical treatment of mixing,
as discussed by \citet{Lattanzio2015}. Further analysis of $^{12}$C/$^{13}$C values
in other clusters observed by APOGEE will allow for a more through mapping of observed
carbon isotope ratios across a range of RGB masses and metallicities, which can be
compared to models of thermohaline mixing.

A more precise determination of the specific evolutionary status for red giants can be made through asteroseismology. NGC~6791 is part of the Kepler field and has extensive asteroseismic information available. \citet{Pinsonneault2014} combined Kepler photometric data with APOGEE observation in the APOKASC catalog. This catalog contains the estimated evolutionary status information using the compilation of \citet{Mosser2014}. 
Eight stars from this sample have an identified evolutionary status (Table 1) from the APOKASC catalog, with 6 classified as RGB and 2 as RC.  The other 3 stars are all quite
cool (T$_{\rm eff}$$\sim$ 3500-3800K and are likely to be luminous RGB stars.  It is worth
noting that the mean value of $^{12}$C/$^{12}$C for the 9 likely RGB stars is 9.3, 
compared to 7.1 for the two RC stars.

\begin{figure*}
\centering
\includegraphics[width=4in,angle=270]{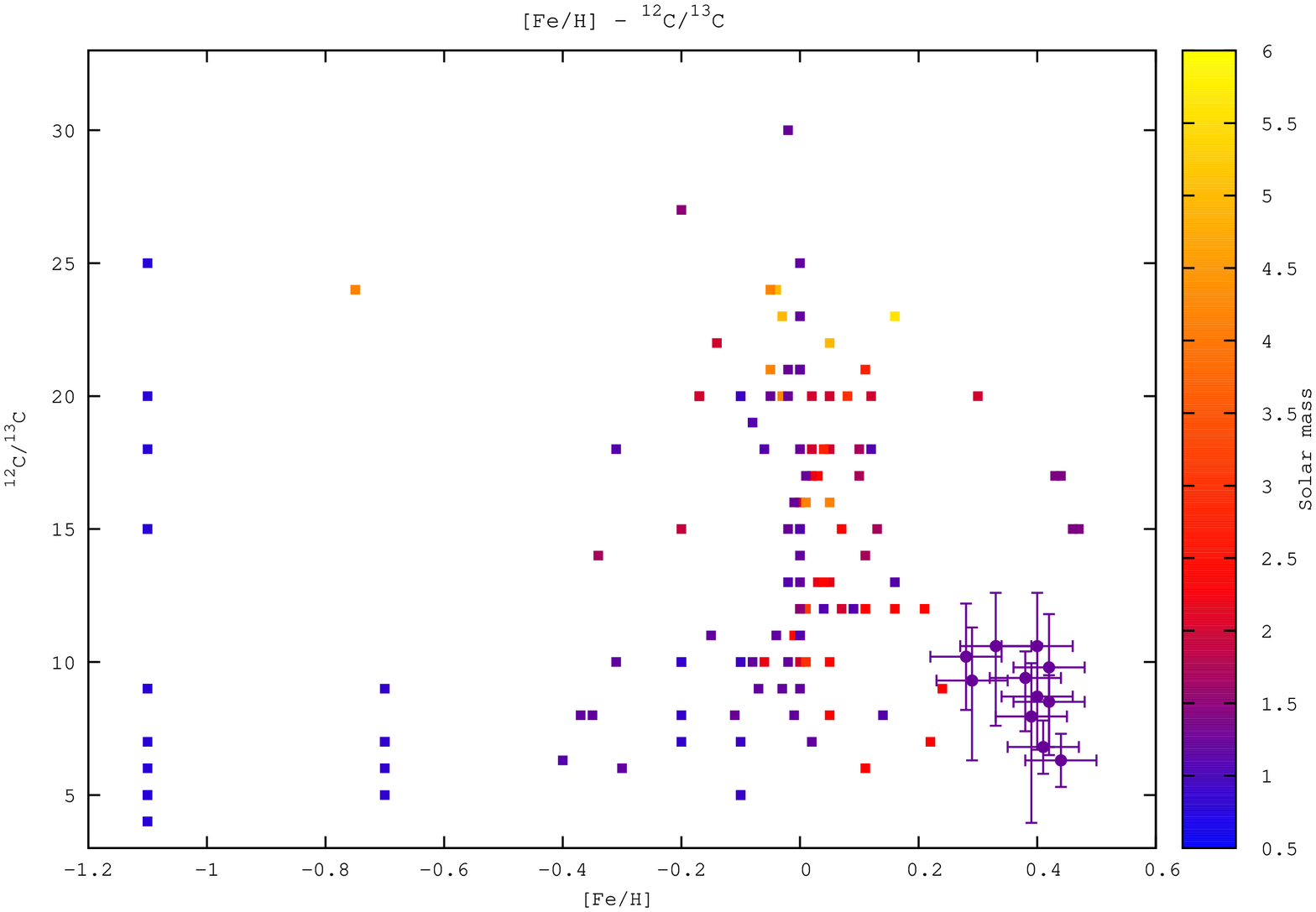}
\label{fig:FeCmassratio}
\caption{ $^{12}$C/$^{13}$C ratio as a function of metallicity from the literature \citep{Gilroy1989, Gilroy1991, Tautvaisiene2000, Shetrone2003_a, Tautvaisiene2005, Smiljanic2009, Mikolaitis2010, Mikolaitis2011_a, Mikolaitis2011_b, Mikolaitis2012, Santric2013,  Tautvaisiene2016, Drazdauskas2016, Dradauskas2016_b, Thompson2010} and this paper 
(circles with error bars). Color-coded with the turn-off mass. We used isochrones \citep{Marigo2017} to determine the turn-off mass of clusters where literature sources were not available.}
\end{figure*}

\subsection{Dependence on metallicity and stellar mass}

Besides evolutionary status, the carbon isotopic ratios depend on stellar mass and perhaps on metallicity, as shown theoretically by \citet{Lagarde2012}. 
Our aim is to collect literature data and directly compare $^{12}$C/$^{13}$C ratios at various metallicities and masses with 
theoretical models. 
Currently, published thermohaline-mixing models take into account stellar mass and metallicity \citep{Lagarde2017}. 
These models cover the initial mass from 0.6 M$_{\odot}$ to 6.0 M$_{\odot}$ with five metallicities [Fe/H] = -2.15, 
-0.54, -0.23, 0 and 0.51. Our new measurements lie in the metal rich region than what was poorly observed before.

\citet{Gilroy1991} found that in M67, a near-solar metallicity cluster, the least evolved faint stars had significantly higher $^{12}$C/$^{13}$C ratios ($>$40), and as stars advanced further in the evolutionary track, the carbon isotopic ratios decreased to about 13. 
While a correlation between $^{12}$C/$^{13}$C with 
T$_{\rm eff}$ could be observed in M67, suggesting that less evolved stars exhibit no extra mixing, the extra mixing in 
NGC~6791 occurs throughout the full extent of the giant branch. We would like to note that our target selection, unlike that of \citet{Gilroy1991}, did not contain stars near the turn-off and sub-giant branch 
(see the right panel of Figure \ref{fig:TCratio}) and also the comparison has to be done carefully because in metal-rich clusters the bump moves down in luminosity. Thus, the mixing mechanism and equilibrium will occur at a different log~g.

Stars with solar-like and slightly metal poor metallicities have been observed and discussed in the literature 
\citep[e.g][]{Gilroy1989, Gilroy1991, Tautvaisiene2000, Shetrone2003_a, Tautvaisiene2005, Smiljanic2009, Mikolaitis2010, Mikolaitis2011_a, Mikolaitis2011_b, Mikolaitis2012, Santric2013,  Tautvaisiene2016, Drazdauskas2016, Dradauskas2016_b}. A collection of these literature results for $^{12}$C/$^{13}$C is plotted in Figure 4 
as a function of metallicity. 
These observed stars cover stellar masses from 1.2 to 5.6 M$_{\odot}$.
Most studies find that above 2.5 solar masses, the carbon isotopic ratio is $\sim$ 22-30, while in lower mass giants, the $^{12}$C/$^{13}$C decreases with mass. 

Comparing results from different literature sources is difficult because evolutionary status information is usually not provided. The 15 red-giant stars from M~67 had similar masses to the ones in NGC~6791, observed by \citet{Gilroy1991}. 
The turn-off mass of the cluster determined from theoretical isochrone fitting was found to be about 1.2 M$_{\odot}$. Thermohaline mixing occurs after the luminosity-bump \citep{Charbonnel2010}, and it is believed that the $^{12}$C/$^{13}$C does not change after the bump. \citet{Gilroy1991} were able to determine the evolutionary status for some of their stars and found that the carbon isotopic ratio for RC stars is between 11 and 13, which is higher than what we find for RC stars in NGC~6791 (around 8).

Comparing results from the literature with our measurements for NGC~6791 is challenging because we do not know the evolutionary status of most of these stars. Based on theoretical models, \citet{Lagarde2012} suggested that slightly stronger mixing ($^{12}$C/$^{13}$C $\sim$ 8.36 ) occurs in more metal-rich stars, which is in line with our measurements. 

Another difficulty is that the isotopic ratios were not derived in a consistent way using the same wavelength regions and analysis method. From these we conclude that this combined dataset is not enough to examine the connection between isotopic ratios, evolutionary status, mass and metallicity in more detail than what was carried out in previous literature studies. Such a study will be possible by determining the $^{12}$C/$^{13}$C ratios of the APOKASC giants, in which accurate masses, metallicities and evolutionary status are available.


\section{Conclusions}

We demonstrated the presence of additional mixing beyond first dredge-up dilution in the atmospheres of RGB/AGB stellar members of the metal-rich open cluster NGC~6791 from measurements of the carbon isotope ratios of $^{12}$C/$^{13}$C. 
The studied stars in NGC~6791 are all beyond the luminosity bump, and our results show good agreement with thermohaline mixing models, if the stars are either on the RC or the AGB. 
Four stars with derived surface gravities between log~g 1 -- 2 appear to be on the AGB based on their isotopic ratios, while stars with values of log~g less than $\sim$1 can either be AGB or RGB, as their evolutionary status is not available. Four stars with log~g $\sim$2.5 appear to be RC stars, and one of them is in fact, confirmed to be on the RC in the APOKASC catalog. 

Carbon isotopic ratios from the literature, combined with our results, do not suggest that extra mixing is stronger at higher 
metallicities, however, the size of the sample is small so far. We cannot conclude that such correlation exists, because we do not know the evolutionary status of most of the stars from the literature. A more detailed study of the effect of mass and metallicity on the isotopic ratios can be only made with the knowledge of the evolutionary status of the stars. 

A methodical study of thermohaline mixing is only possible by observing large number of stars in the context \citep{Lagarde2015} of large sky surveys. APOGEE is such a survey capable of determining values of $^{12}$C/$^{13}$C for thousands of stars. These new measurements will
be available in the $15^{\rm th}$ data release of SDSS in 2018. 

\section*{Acknowledgements} 

L. Sz. and Sz. M. has been supported by the Hungarian 
NKFI Grants K-119517 of the Hungarian National Research, Development and Innovation Office. Sz. M. has been supported by the Premium Postdoctoral Research Program of the Hungarian Academy of Sciences.

J.G.F-T gratefully acknowledge support from the Chilean BASAL Centro de Excelencia an Astrof{\'i}sica Technolog{\'i}as Afines (CATA) grant PFB-06/2007

S.V. gratefully acknowledges the support provided by Fondecyt reg. n. 1170518

DAGH was funded by the Ram{\'o}n y Cajal fellowship number RYC-2013-14182 and he acknowledges support provided by
the Spanish Ministry of Economy and Competitiveness (MINECO) under grant AYA-2014-58082-P.

Funding for the Sloan Digital Sky Survey IV has been provided by the
Alfred P. Sloan Foundation, the U.S. Department of Energy Office of
Science, and the Participating Institutions. SDSS acknowledges
support and resources from the Center for High-Performance Computing at
the University of Utah. The SDSS web site is www.sdss.org.

SDSS is managed by the Astrophysical Research Consortium for the Participating Institutions of the SDSS Collaboration including the Brazilian Participation Group, the Carnegie Institution for Science, Carnegie Mellon University, the Chilean Participation Group, the French Participation Group, Harvard-Smithsonian Center for Astrophysics, Instituto de Astrof{\'i}sica de Canarias, The Johns Hopkins University, Kavli Institute for the Physics and Mathematics of the Universe (IPMU) / University of Tokyo, Lawrence Berkeley National Laboratory, Leibniz Institut f{\"u}r Astrophysik Potsdam (AIP), Max-Planck-Institut f{\"u}r Astronomie (MPIA Heidelberg), Max-Planck-Institut f{\"u}r Astrophysik (MPA Garching), Max-Planck-Institut f{\"u}r Extraterrestrische Physik (MPE), National Astronomical Observatories of China, New Mexico State University, New York University, University of Notre Dame, Observat{\'o}rio Nacional / MCTI, The Ohio State University, Pennsylvania State University, Shanghai Astronomical Observatory, United Kingdom Participation Group, Universidad Nacional Aut{\'o}noma de M{\'e}xico, University of Arizona, University of Colorado Boulder, University of Oxford, University of Portsmouth, University of Utah, University of Virginia, University of Washington, University of Wisconsin, Vanderbilt University, and Yale University.









\bsp	
\label{lastpage}
\end{document}